\documentclass[sigconf]{acmart}
\usepackage{tcolorbox}

\usepackage{fancyhdr}
\AtBeginDocument{%
  \fancyhead{} % 清空页眉（左、右、居中）
  \fancyfoot[C]{\thepage} % 保留页码在页脚居中
  \thispagestyle{plain} % 标题页使用 plain 样式（无页眉，有页码）
}

\makeatletter
\let\oldmaketitle\maketitle
\renewcommand{\maketitle}{%
  \oldmaketitle%
  \thispagestyle{plain}% Ensure title page uses plain style
}
\makeatother

\AtBeginDocument{%
  }

\renewcommand\footnotetextcopyrightpermission[1]{}
\setcopyright{acmlicensed}
\copyrightyear{2018}
\acmYear{2018}
\acmDOI{XXXXXXX.XXXXXXX}
\acmConference[Conference acronym 'XX]{Make sure to enter the correct
  conference title from your rights confirmation email}{June 03--05,
  2018}{Woodstock, NY}
\acmISBN{978-1-4503-XXXX-X/2018/06}

\begin{document}

\title{Mind2Matter: Creating 3D Models from EEG Signals}

\author{Xia Deng}
\authornote{Both authors contributed equally to this research.}
\affiliation{%
  \institution{East China University of Science and Technology}
  % \city{Shanghai}
  \country{}}
  \email{xiadeng@mail.ecust.edu.cn}

\author{Shen Chen}
\authornotemark[1]
\affiliation{%
  \institution{East China University of Science and Technology}
  % \city{Shanghai}
  \country{}}
  \email{shenchen33@mail.ecust.edu.cn}

\author{Jiale Zhou}
\authornote{Corresponding author.}
\affiliation{%
  \institution{East China University of Science and Technology}
  % \city{Shanghai}
  \country{}}
  \email{zhou.jiale@ecust.edu.cn}

\author{Lei Li}
\authornotemark[2]
\affiliation{%
  \institution{University of Copenhagen, University of Washington}
  % \city{Seattle}
  % \state{Washington}
  \country{}}
  \email{lilei@di.ku.dk}

\begin{abstract}
 The reconstruction of 3D objects from brain signals has gained significant attention in brain-computer interface (BCI) research. Current research predominantly utilizes functional magnetic resonance imaging (fMRI) for 3D reconstruction tasks due to its excellent spatial resolution. Nevertheless, the clinical utility of fMRI is limited by its prohibitive costs and inability to support real-time operations. In comparison, electroencephalography (EEG) presents distinct advantages as an affordable, non-invasive, and mobile solution for real-time brain-computer interaction systems. While recent advances in deep learning have enabled remarkable progress in image generation from neural data, decoding EEG signals into structured 3D representations remains largely unexplored. In this paper, we propose a novel framework that translates EEG recordings into 3D object reconstructions by leveraging neural decoding techniques and generative models. Our approach involves training an EEG encoder to extract spatiotemporal visual features, fine-tuning a large language model to interpret these features into descriptive multimodal outputs, and leveraging generative 3D Gaussians with layout-guided control to synthesize the final 3D structures. Experiments demonstrate that our model captures salient geometric and semantic features, paving the way for applications in brain-computer interfaces (BCIs), virtual reality, and neuroprosthetics. Our code is available in \url{https://github.com/sddwwww/Mind2Matter}. 
\end{abstract}

\begin{CCSXML}
<ccs2012>
   <concept>
       <concept_id>10010147.10010371</concept_id>
       <concept_desc>Computing methodologies~Computer graphics</concept_desc>
       <concept_significance>500</concept_significance>
       </concept>
   <concept>
       <concept_id>10003120.10003121</concept_id>
       <concept_desc>Human-centered computing~Human computer interaction (HCI)</concept_desc>
       <concept_significance>500</concept_significance>
       </concept>
 </ccs2012>
\end{CCSXML}

\ccsdesc[500]{Computing methodologies~Computer graphics}
\ccsdesc[500]{Human-centered computing~Human computer interaction (HCI)}

\keywords{Electroencephalogram, LLM, 3D gaussian splatting}
\begin{teaserfigure}
\centering
  \includegraphics[width=0.7\textwidth]{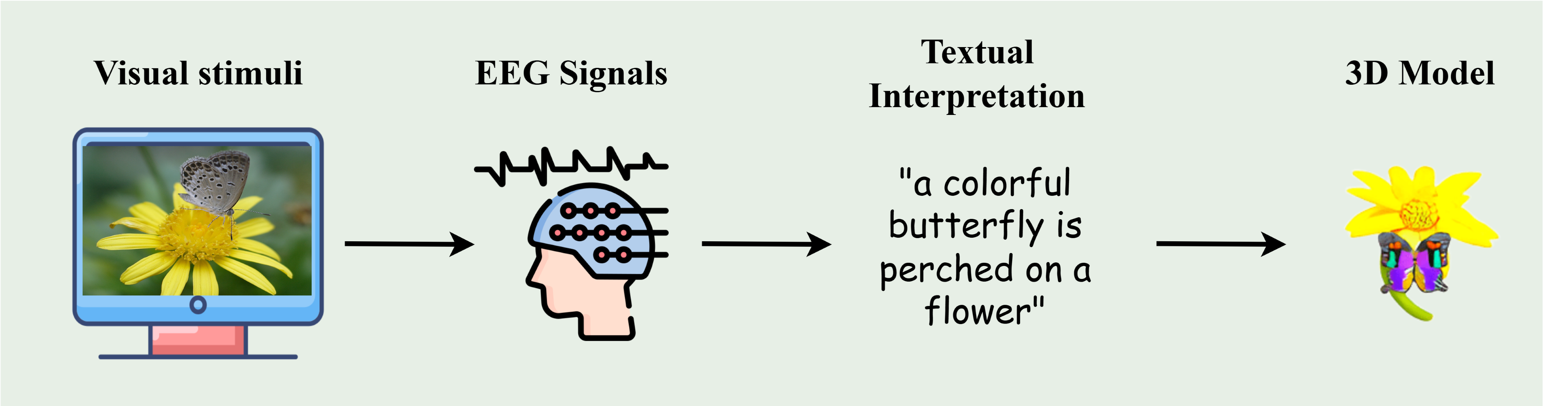}
  \caption{Illustration from visual stimuli to 3D model generation. A subject receives a visual input(left), and EEG signals are recorded. These signals are processed, interpreted as text (middle), and used to generate a 3D model of the scene (right), translating brain activity into a visual representation.}
  \Description{3D object reconstruction using EEG signals is depicted.}
  \label{fig:teaser}
\end{teaserfigure}

\maketitle

\section{Introduction}
Decoding brain signals to reconstruct 3D objects represents a transformative frontier in brain-computer interface (BCI) research \cite{brumberg2018brain,pandarinath2017high}, with profound implications for human-computer interaction, neuroscience, and assistive technologies. By translating neural activity into tangible 3D representations, such advancements could enable intuitive control of virtual environments, facilitate neuroprosthetic design \cite{borton2013personalized}, and deepen our understanding of how the brain encodes complex visual perceptions \cite{naselaris2009bayesian,horikawa2017generic}. Unlike traditional input methods that rely on manual or verbal commands, brain signal decoding offers a direct, non-invasive pathway to bridge the gap between mental imagery and digital output. This capability is particularly valuable for applications requiring real-time interaction, such as immersive virtual reality \cite{slater2016enhancing} and creative 3D modeling \cite{nishino20013d}, where seamless integration of human intent and computational systems is paramount.

Despite the promise of neural decoding, existing research has predominantly focused on reconstructing perceptual experiences using functional magnetic resonance imaging (fMRI). Leveraging fMRI’s high spatial resolution, many studies have successfully mapped brain activity to visual reconstructions \cite{takagi2023high,chen2023seeing,xia2024dream,scotti2023reconstructing,gao2024mind,gao2025mind3dadvancingfmribased3d}. However, the clinical utility of fMRI remains substantially limited by its prohibitive cost, lack of portability, and insufficient temporal resolution to support real-time operation \cite{lan2023seeing}. In contrast, electroencephalography (EEG) represents a cost-effective, portable, and non-invasive neuroimaging modality that offers distinct advantages for visual reconstruction and real-time applications \cite{heckenlively2006principles}.

This study pioneers a novel approach for reconstructing three-dimensional objects from EEG signals evoked by visual stimuli, utilizing 3D Gaussian splatting \cite{kerbl20233d} to bridge the gap between neural activity and structured volumetric representations. The endeavor confronts substantial methodological challenges stemming from fundamental limitations of scalp-recorded EEG. Unlike fMRI, EEG struggles with low spatial resolution and a low signal-to-noise ratio, which complicates the extraction of robust neural representations. Additionally, EEG signals often fail to effectively capture high-level semantic information, and the complexity of generative models further hinders precise learning of structural details \cite{hebb2005organization}. While EEG-based decoding has made strides in 2D image generation \cite{kavasidis2017brain2image,mishra2023neurogan,bai2023dreamdiffusion,singh2024learning,pan2024reconstructing}, the extension to veridical 3D object recovery remains largely unexplored territory.

To overcome these challenge, we propose Mind2Matter, a novel two-stage framework that bridges neural decoding and 3D generation through hierarchical semantic understanding. In the first stage, we train a deep multi-channel neural encoder to capture localized semantic features from EEG recordings. This encoder effectively extracts spatiotemporal patterns, which are then projected into a labeled embedding space to fine-tune a language model. By aligning EEG embeddings with semantic representations, the fine-tuned language model generates detailed textual descriptions of the observed visual stimuli, such as object identities and their spatial relationships. This cross-modal translation leverages the language model's inherent world knowledge \cite{cai2025bayesian,li2024cpseg} to compensate for EEG's semantic sparsity, effectively converting noisy neural patterns into structured object descriptions.

While most methods focus on reconstructing single objects, human visual perception is intrinsically holistic, simultaneously encoding multiple objects along with their functional properties and spatial configurations. In the second stage of Mind2Matter, we leverage a scene-level text-to-3D framework based on generative 3D Gaussian techniques \cite{chen2024scalinggaussian,liu2024graph,yan20243dsceneeditor} to synthesize complex 3D scenes. This framework takes the generated textual descriptions as input and produces high-fidelity, coherent 3D reconstructions featuring multiple objects with precise interactions. By incorporating layout-guided control, our approach ensures that the synthesized 3D scenes accurately reflect the spatial and semantic relationships described in the text, overcoming the limitations of EEG’s low spatial resolution. Through this two-stage pipeline, Mind2Matter not only demonstrates the feasibility of EEG-based 3D reconstruction but also establishes a scalable pathway for real-time BCI applications, with potential impacts in virtual reality, neuroprosthetics, and intuitive 3D design.

In summary, this paper makes the following contributions:
\begin{itemize}
\item We presents Mind2Matter, a novel framework for EEG-to-3D scene reconstruction, which enables the extraction of semantic and spatial features from EEG signals to reconstruct complex, multi-object 3D scenes.
\item A two-stage neural encoding-decoding framework is proposed, facilitating the implicit encoding of EEG signals into textual descriptions and the subsequent decoding into coherent 3D scenes with precise object interactions.
\item Experimental results demonstrate that the proposed method can effectively reconstruct complex 3D scenes with multiple objects using EEG signals, highlighting its potential for real-time BCI applications.
\end{itemize}

\section{Related Work}
\subsection{Generating texts from brain activities}

The generation of textual descriptions from brain activities has emerged as a promising direction in brain-computer interface (BCI) research, aiming to translate neural signals into interpretable semantic outputs. Early research \cite{biswal2019eegtotext,srivastava2020think2type,liu2024eeg2text} predominantly focused on closed-vocabulary approaches, wherein the decoding process was restricted to a predefined set of words or phrases. Notably, Think2type \cite{srivastava2020think2type} converted activity intentions of disabled individuals into Morse code representations, which were then transformed into corresponding text. While these methods achieved initial success in controlled environments, they encountered limitations regarding scalability and addressing the complexity of natural language. More recent studies \cite{feng2023aligning,amrani2024deep,tao2025see,chen2025decoding} have transcended vocabulary constraints through the development of open-vocabulary text generation systems. These approaches utilize deep learning architectures and language models to generate unrestricted textual outputs from neural signals. These investigations addressed several critical challenges, including individual variability in EEG patterns \cite{amrani2024deep}, cross-modal representation learning \cite{tao2025see}, and capturing complex temporal dependencies in neural data \cite{chen2025decoding}.

To address challenges such as defining word-level boundaries in EEG signals and other language processing tasks, several investigations \cite{mishra2024thought2text,ikegawa2024text} have proposed language-agnostic solutions that capture signals through image modality and leverage advancements in image-text intermodality to generate text from the collected data. Despite these advancements, challenges persist in achieving precise neural-linguistic mapping and generating contextually appropriate descriptions. Novel architectures and training strategies need to be explored to enhance the accuracy and naturalness of EEG-to-text conversion.

\subsection{Text-Driven 3D Generation}

Text-driven 3D generation has become a rapidly evolving field in computer vision and graphics, aiming to synthesize 3D objects or scenes directly from textual descriptions. Early methods \cite{jain2022zero,mohammad2022clip} using CLIP \cite{radford2021learning} for optimization often yielded inconsistent results but lacked realism and fidelity. Recent approaches leverage 2D diffusion models, such as Zero-1-to-3 \cite{liu2023zero} and MVDream \cite{shi2023mvdream}, to generate 3D assets from multi-view images. 

Another line of research \cite{lin2023magic3d,chen2023control3d,zhang2024text2nerf} has focused on NeRF-based text-to-3D generation, utilizing neural radiance fields (NeRF) \cite{mildenhall2021nerf} to create 3D representations from text. Methods like DreamFusion \cite{poole2022dreamfusion} introduce score distillation sampling (SDS) to optimize NeRF parameters using a pre-trained 2D diffusion model, achieving high-quality single-object reconstructions. However, NeRF-based approaches suffer from low efficiency due to their computationally intensive optimization process. To address this, recent works \cite{chen2024text,wu2024consistent3d,zhang2024gaussiancube,liang2024luciddreamer} have combined diffusion models with 3D Gaussian splatting (3DGS) to achieve rapid and high-fidelity model generation. 

\begin{figure*}[t]
  \centering
  \includegraphics[width=0.9\textwidth]{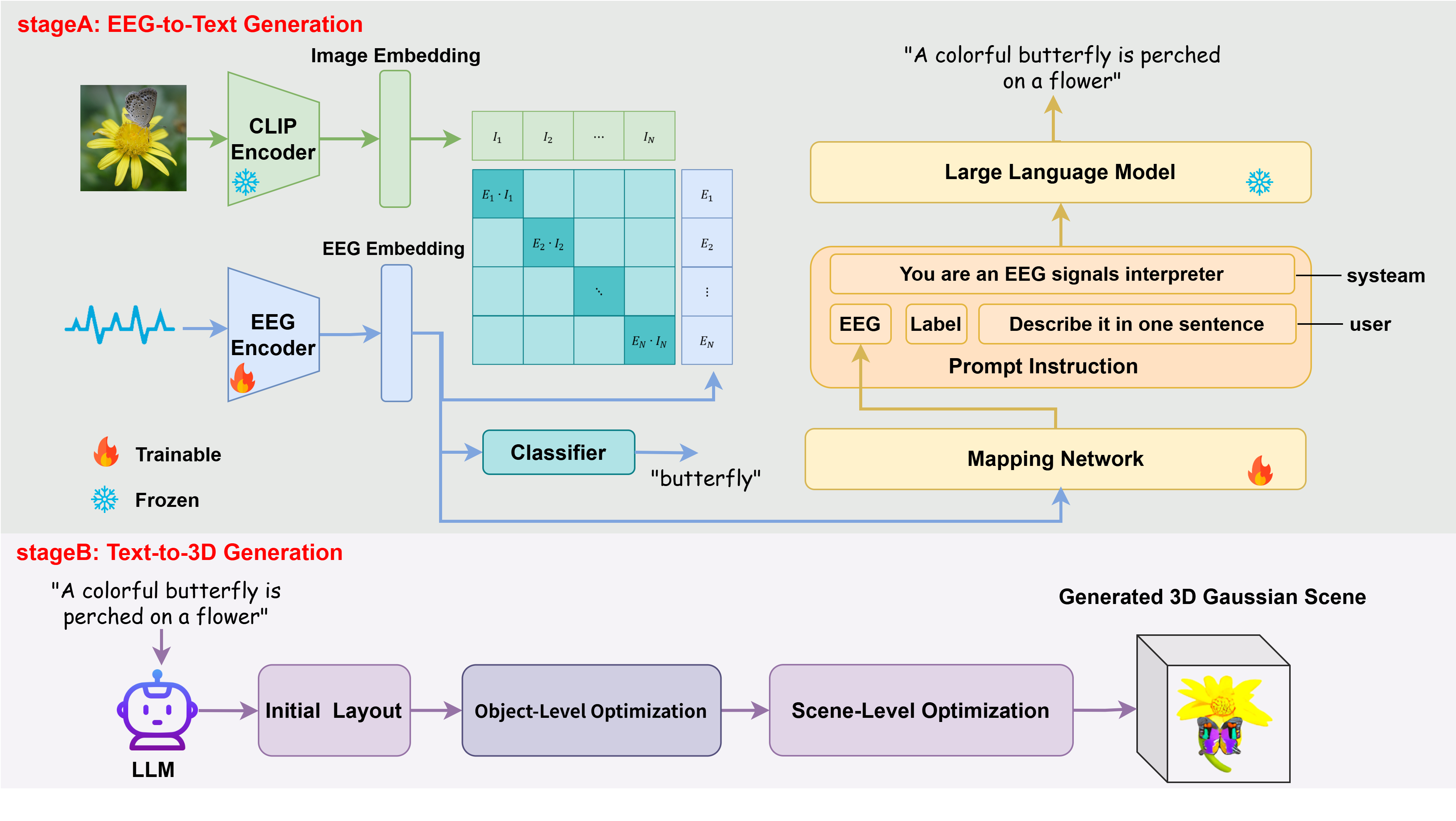} 
  \caption{Architecture of Mind2Matter. EEG signals are processed by a trainable EEG Encoder to extract spatiotemporal features, generating EEG embeddings aligned with image embeddings from a frozen CLIP Encoder. These embeddings are transformed by a trainable Mapping Network and fed into a frozen LLM, which generates a textual description (e.g., "A colorful butterfly is perched on a flower") using a prompt. The text is then used by another LLM to create an initial 3D layout, followed by object-level and scene-level optimization with 3D Gaussian splatting and diffusion priors, producing a high-fidelity 3D scene.}
  \Description{}
  \label{pipline}
\end{figure*}

\section{Methods}
As shown in Fig.\ref{pipline}, our method employs a two-stage pipeline to reconstruct 3D scenes from EEG signals, tackling their low spatial resolution and semantic sparsity. A multi-hierarchical encoder and fine-tuned language model first generate textual descriptions via cross-modal alignment, which are then converted into 3D scenes using layout-constrained 3D Gaussian splatting. Details follow in subsequent subsections.
\subsubsection*{\textbf{Preliminaries}}
3D Gaussian Splatting (3DGS) \cite{kerbl20233d} is a recent advancement in 3D scene representation and rendering, offering an efficient alternative to neural radiance fields (NeRF) \cite{mildenhall2021nerf} for real-time applications. 3DGS represents a scene as a set of 3D Gaussians, denoted as $\{G_n|n = 1,2,\ldots N\}$. Each Gaussian $G_n$ is parameterized by a position $\mu_n\in\mathbb{R}^3$, a covariance matrix $\Sigma_n\in\mathbb{R}^3$, a color $c_n\in\mathbb{R}^3$, and an opacity $\alpha_n\in\mathbb{R}$. The Gaussian function is defined as:
\begin{equation}
G_n(p) = \exp\left(-\frac{1}{2}(p - \mu_n)^T \Sigma_n^{-1} (p - \mu_n)\right)
\end{equation}
Where $\Sigma_n$ is parameterized by a rotation matrix $R^n\in\mathbb{R}^4$ and a scaling matrix $S^n\in\mathbb{R}^3$ to ensure positive definiteness, such that $\Sigma^n = R^nS^nS_n^TR_n^T$.

For rendering, 3DGS projects these Gaussians onto 2D camera planes using differentiable splatting. Given a viewing transformation $W_n$ and Jacobian $J_n$ of the projective transformation, the covariance in camera coordinates is computed as $\Sigma_n' = J_nW_n\Sigma^nW_n^TJ_n^T$. The color along a ray $\mathbf{r}$ is then calculated as:
\begin{equation}
C(\mathbf{r}) = \sum_{i\in N} c_i\alpha_i \prod_{j=1}^{i-1} (1 - \alpha_j), \quad \alpha_i = \alpha_iG_i^{2D}(p)
\end{equation}
where $N$ denotes the number of Gaussians along the ray, and $G_i^{2D}(p)$ is the 2D projection of the Gaussian.

\subsection{EEG-to-Text Generation}
\subsubsection*{\textbf{EEG Encoder}}
The EEG encoder constitutes a foundational framework for decoding neural signals, with its primary function being the extraction of multi-hierarchical semantic features from raw EEG data. We introduce a unified encoder architecture that processes input signals via a modular, hierarchically structured composition. 

\begin{figure}[t]
  \centering
  \includegraphics[width=0.9\linewidth]{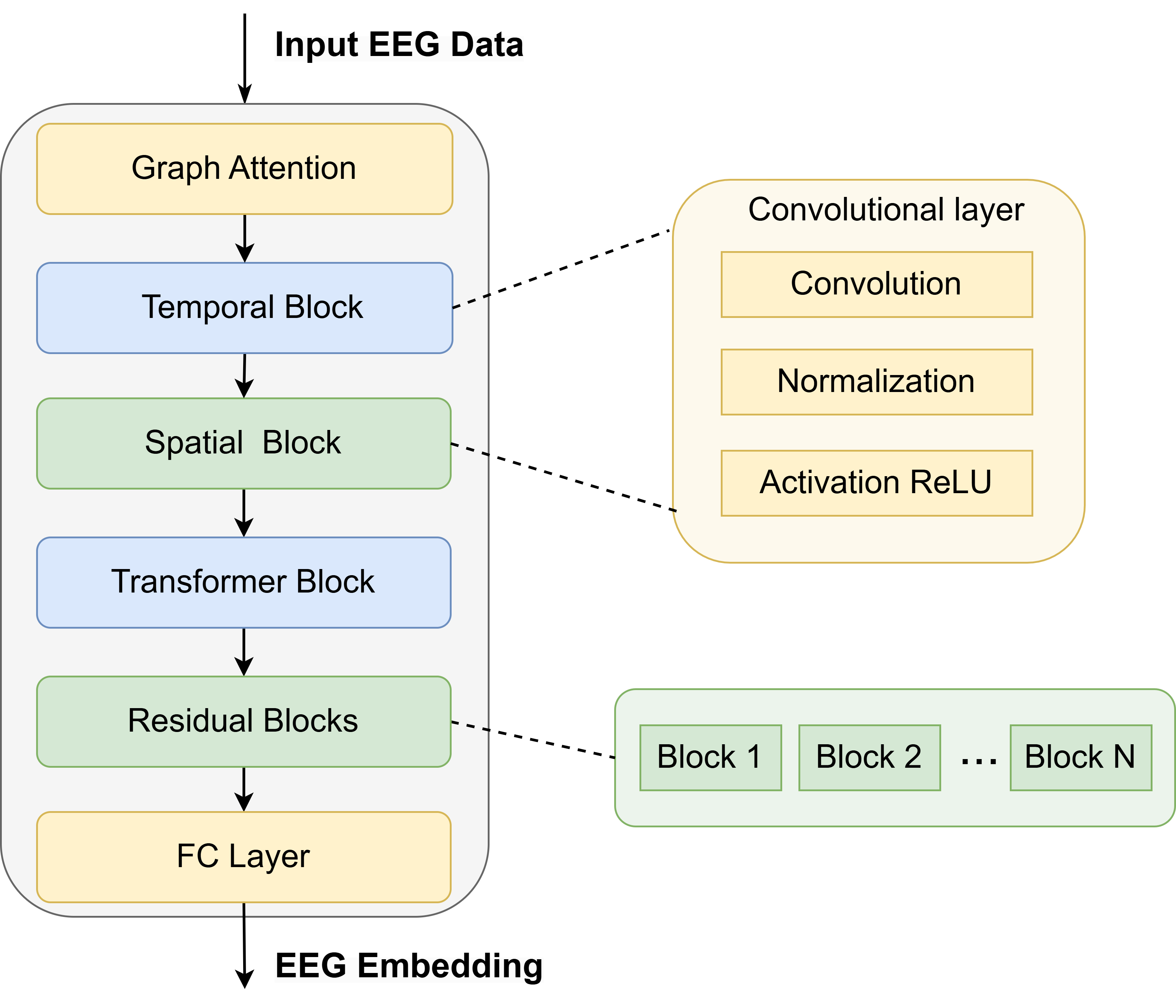}
  \caption{Architecture of the EEG Encoder}
  \Description{ }
  \label{EEG_Encoder}
\end{figure}

As illustrated in Fig.\ref{EEG_Encoder}, the proposed encoder begins with a Graph Attention Module \cite{brody2021attentive}, which leverages graph attention mechanisms to model the relational dependencies among EEG electrodes. This block constructs a graph representation of the EEG channels, where nodes represent electrodes and edges capture their interactions, enabling the encoder to focus on relevant inter-electrode relationships. To enhance the extraction of temporal features, we adopt multi-scale temporal convolution, using a series of two-dimensional convolution modules with gradually increasing dilation rates. This design expands the receptive field without sacrificing the resolution, allowing the encoder to capture long-range dependencies at different time scales. Subsequently, spatial convolution is utilized to further optimize the spatial features by exploiting inter-electrode relationships. Moreover, we augment the architecture with a Transformer-based module to model long-range dependencies inherent in EEG signals. Inspired by residual learning principles, we incorporate residual blocks with bottleneck designs to ensure efficient feature propagation and gradient stability during training. The encoder architecture culminates in a final convolutional layer that integrates and refines extracted features, generating a compact and semantically rich representation of the input EEG signal. For implementation details, each convolutional block consists of a convolutional layer, followed by normalization and ReLU activation, structured as a series of modular units. 

\subsubsection*{\textbf{Cross-modal Alignment}}
The model is trained by minimizing two distinct loss functions: 
\begin{enumerate}
    \item a classification cross-entropy loss (CE) between the predicted object labels and the ground truth ImageNet labels.
    \item the Contrastive-Adaptive Margin Loss (CAML) between the EEG embeddings and the image embeddings derived from the pre-trained CLIP model.
\end{enumerate}
These two losses jointly optimize the cross-modal alignment and the classification task.
% (1) a classification cross-entropy loss (CE) between the predicted object labels and the ground truth ImageNet labels, and (2) the Contrastive-Adaptive Margin Loss (CAML) between the EEG embeddings and the image embeddings derived from the pre-trained CLIP model. 

The classification objective is formulated as a cross-entropy loss between the predicted object labels and ground-truth labels:
\begin{equation}
\mathcal{L}_{CE} = -\sum y_i log(p_i),
\end{equation}
where $y_i$ is the ground-truth label and $p_i$ is the predicted probability distribution over object categories obtained from the classifier.

The conventional contrastive loss approaches employ fixed margin thresholds, which fail to adapt to varying semantic similarities between different sample pairs. Inspired by the margin adaptation mechanisms in human perceptual learning \cite{dosher1998perceptual}, we propose an adaptive margin contrast loss (AMCL) that automatically adjusts optimization boundaries based on sample-wise EEG embedding characteristics. The ordinary contrastive learning uses InfoNCE loss\cite{oord2018representation}:
\begin{equation}
\mathcal{L}_{InfoNCE} = -\frac{1}{N}\sum_{i=1}^N log \frac{exp(S_{E,I}/\tau)}{\sum_{k=1}^N exp(S_{E,I_k}/\tau)}, 
\end{equation}
where the $S_{E,I}$ denotes the similarity score between EEG signal $E$ and image $I$ pairing data, and $\tau$ is a temperature parameter. The InfoNCE loss encourages paired samples to be closer in the embedding space while pushing unpaired samples apart.

To adaptively adjust the discrimination boundary between positive and negative pairs, we introduce an adaptive margin mechanism. For each sample pair, the margin $m_i$ is dynamically computed based on their similarity:
\begin{equation}
m_i = \alpha \cdot (1 - S_{E,I}),
\end{equation}
where $\alpha$ is a scaling factor that controls the margin's sensitivity to similarity. The adaptive margin loss is then defined as:
\begin{equation}
\mathcal{L}_{margin} = \frac{1}{N}\sum_{i=1}^N \frac{1}{k}\sum_{j\neq i} max(0, m_i - (S_{E,i} - S_{E,I_k})).
\end{equation}

The overall training objective can be written as:
\begin{equation}
\mathcal{L}_{total} = \mathcal{L}_{CE} + \lambda_1\mathcal{L}_{InfoNCE} + \lambda_2\mathcal{L}_{margin},
\end{equation}
where $\lambda_1$, $\lambda_2$ is a balancing coefficient that controls the relative importance of the cross-modal alignment term. This joint optimization enables our model to simultaneously learn discriminative EEG representations for object recognition and establish robust alignment between EEG and visual embeddings.

\subsubsection*{\textbf{Language model fine-tuning}}
During training, a primary problem is the transformation between EEG embeddings and language model representations. Consequently, we initially proposed to fine-tune the language model while training the mapping network. This approach provided additional flexibility to the network and yielded more expressive results. However, directly fine-tuning pre-trained language models introduces a substantial number of trainable parameters. \cite{li2021prefix} proposed prefix-tuning as a lightweight alternative to fine-tuning for natural language generation (NLG) tasks, training an upstream prefix that guides the downstream LM while keeping the latter unchanged. Inspired by this concept, we propose to avoid fine-tuning and maintain the language model fixed during training, only training the mapping network, thereby achieving a more lightweight model.

The mapping network serves as a crucial interface, transforming EEG embeddings into a format compatible with the LLM. We implement a multi-layer perceptron (MLP)-based mapping network, with the mapping process defined as:
\begin{equation}
H_{hidden} = ReLU(W_1H_{eeg} + b_1)
\end{equation}
\begin{equation}
H_i = W_2 \cdot H_{hidden} + b_2, \quad \text{for } i = 1,2,\ldots, N
\end{equation}
where $W_1$, $W_2$, $b_1$ and $b_2$ are learnable parameters of the mapping network. Once transformed, these embeddings are then concatenated with token embeddings derived from the input prompt, enabling the LLM to seamlessly process both text and EEG embeddings.

The structure of the input prompt incorporates language instructions and actual ImageNet class labels, formatted as follows:
\begin{tcolorbox}[colback=gray!10,colframe=white]
\textbf{system}: You are an EEG signal interpreter.

\textbf{user}: <EEG><Label> Describe it in one sentence.
\end{tcolorbox}
\noindent where the <EEG> token is replaced by the mapped EEG embeddings, and the <Label> token is replaced by the ground truth ImageNet class label. The tokenized prompt, incorporating EEG-derived embeddings, is concatenated with special tokens to form the input sequence for the LLM.

The fine-tuning objective is to minimize the cross-entropy loss between the LLM's predicted text and the ground truth description derived from the ImageNet class label. Formally, let $\mathcal{T}$ denote the tokenized prompt, and $Y$ represent the ground truth description. The LLM, parameterized by $\theta_{LLM}$, generates a predicted sequence $\hat{Y}$ conditioned on the input and $\mathcal{T}$. The fine-tuning loss is defined as:
\begin{equation}
\mathcal{L}_{LLM} = -\frac{1}{M}\sum_{m=1}^M \log P(Y_m|\{H_1, H_2,\ldots,H_N\}, \mathcal{T}; \theta_{LLM}),
\end{equation}
where M is the length of the ground truth description sequence, and P($\cdot$) is the probability assigned by the LLM to the correct token at each position. Through the utilization of MLP-based mapping network, the LLM is enabled to generate accurate and contextually relevant descriptions of visual stimuli based on brain electrical signals, thereby establishing a bridge between neural activity and language comprehension.

\subsection{Text-to-3D Generation}
Having successfully generated textual descriptions from EEG signals in Stage A, Stage B focuses on transforming these text outputs into 3D object representations, completing the pipeline from brain signals to visual reconstructions. Initially, LLMs interpret the EEG-derived text to produce an initial layout of the 3D scene, capturing the spatial arrangement of objects as conceptualized from brain activity. Subsequently, a layout-constrained 3D Gaussian optimization constructs the scene using Gaussian splatting, ensuring geometric and textural accuracy under adaptive constraints. This stage not only bridges the gap between neural signals and tangible 3D outputs, but also ensures the reconstructed objects reflect the user’s mental imagery with precision and realism. 

\subsubsection*{\textbf{Generate Initial Layout with LLMs}}
To initialize the scene structure, we employ a Large Language Model (LLM) to directly predict the spatial attributes of all objects from a single input text. For each object entity identified in the prompt, the LLM estimates a corresponding 3D bounding box defined by its center coordinates, size, and orientation.

Formally, given an input description $T$, the LLM produces a set of object-level parameters $\mathcal{B} = \{(x_i, y_i, z_i), (l_i, w_i, h_i), \theta_i\}_{i=1}^N$, where:
\begin{itemize}
    \item $(x_i, y_i, z_i)$ denotes the 3D center of object $o_i$ in the global scene space;
    \item $(l_i, w_i, h_i)$ specifies the size (length, width, height) of the bounding box;
    \item $\theta_i$ represents the yaw-angle of the object with respect to the world frame.
\end{itemize}

The predicted boxes form a preliminary scene layout $\mathcal{B}$, providing a coherent and semantically aligned spatial configuration. This layout serves as a strong initialization for geometry generation, enabling efficient and controllable scene construction without requiring manual placement or handcrafted rules.

\subsubsection*{\textbf{Layout-Constrained 3D Gaussian Optimization}}

With the initial layout provided by the LLM, we proceed to construct the full 3D scene using Gaussian splatting. Each object is modeled as a set of anisotropic 3D Gaussians, with parameters encoding its spatial location, shape, appearance, and visibility. The layout acts as a spatial prior, guiding the initial placement and scale of each object, which is further refined through a two-stage optimization process: first at the object level and then at the scene level.

\paragraph{Object-Level Optimization}

For each object $o_i$, we initialize its Gaussian representation $\mathcal{G}_i$ based on the predicted bounding box and refine it via Score Distillation Sampling (SDS). Leveraging text-conditioned diffusion models\cite{shi2023mvdream,guan2025learning}, we update object parameters by minimizing the SDS gradient:

\begin{equation}
    \nabla_{\mathcal{G}_i} \mathcal{L}_{\text{obj}} = \mathbb{E}_{\epsilon, \xi} \left[ \lambda_{i} \cdot \left( \epsilon_{\theta}\left(I_i^{\text{pred}}, \xi \right) - \epsilon \right) \frac{\partial I_i^{\text{pred}}}{\partial \mathcal{G}_{i}} \right],
\end{equation}
where $I_i^{\text{pred}}$ is the rendered image of object $o_i$, $\epsilon$ is the noise sample, $\xi$ is the diffusion condition (e.g., timestep, view), and $\epsilon_\theta$ is the denoising model. 

\paragraph{Scene-Level Optimization}

To ensure global spatial coherence and semantic alignment, we jointly optimize all objects in the scene using a layout-constrained representation. Each object $i$ is described by its Gaussian parameters $\mathcal{G}_i$, and layout $\mathcal{B}_i$. These components collectively form the scene layout $\mathcal{X}_{\text{scene}} = \left\{ \mathcal{G}_i, \mathcal{B}_i \right\}_{i=1}^N.$

We directly input this layout into the SDS guided by ControlNet and optimize the entire scene:
\begin{equation}
    \nabla_{\mathcal{G}} \mathcal{L}_{\text{sc}} = \lambda \cdot \mathbb{E}_{\epsilon, \psi} \left[ \lambda \cdot \left( \epsilon_\alpha\left(I_{\mathcal{X}_{\text{scene}}}^{\text{pred}}, \psi\right) - \epsilon \right) \frac{\partial I_{\mathcal{X}_{i}}^{\text{pred}}}{\partial \mathcal{G}} \right],
\end{equation}
where $\psi$ denotes the prompt and rendering conditions, and $\epsilon_\alpha$ is the denoising network. This unified optimization updates all spatial and appearance attributes jointly, producing a coherent and physically plausible 3D scene.

% \subsubsection*{\textbf{Refine with Diffusion and Layout}}
% The final stage iteratively optimizes both object geometries and their spatial arrangements using diffusion priors. Multi-view consistency is achieved through a dual optimization strategy: instance-level enhancement with text-aligned diffusion models, followed by global scene harmonization conditioned on layout renders. This process dynamically adjusts initial layout predictions to align with physically plausible 3D structures.

\begin{table*}
\centering
\caption{Quantitative comparison of EEG-to-text generation performance across different EEG encoders. Results are reported in ROUGE-1 (Recall/R, Precision/P, F1-score/F), BLEU (B-1 to B-4), and BERTScore (R/P/F) metrics. Bold values indicate the best performance.}
\label{quetg}
\begin{tabular}{c c c c c c c c c c c}
\toprule
% \textbf{Model} & \textbf{R-1} & \textbf{R-2} & \textbf{R-L} & \textbf{B-1} & \textbf{B-2} & \textbf{B-3} & \textbf{B-4} & \textbf{WER (\%)$\downarrow$} \\
& \multicolumn{3}{c}{\textbf{ROUGE-1(\%)$\uparrow$}} & \multicolumn{4}{c}{\textbf{BLEU(\%)$\uparrow$}} & 
\multicolumn{3}{c}{\textbf{BERT(\%)$\uparrow$}}\\
{\textbf{Model}} & \textbf{R} & \textbf{P} & \textbf{F} & \textbf{B-1} & \textbf{B-2} & \textbf{B-3} & \textbf{B-4} &
\textbf{R} & \textbf{P} & \textbf{F}\\
\midrule
\textbf{Conformer} &30.98 & 34.96 & 31.84 & 31.96 & 15.32 & 9.02 & 6.14 & 34.98 & 35.86 & 35.41 \\
\textbf{EEGNet} & 27.87 & 30.67 & 28.41 & 29.51 & 11.96 & 6.21 & 4.2 & 30.65 & 30.04 & 30.37\\
\textbf{DeepNet} & {23.92} & {27.17} & {24.8} & {27.41} & {9.97} & {4.79} & {3.19} & {26.81} & {28.33} & {27.6}\\
\textbf{Ours} & \textbf{33.12} & \textbf{37.38} & \textbf{34.21} & \textbf{34.33} & \textbf{17.31} & \textbf{10.6} & \textbf{7.62} & \textbf{36.84} & \textbf{37.54} & \textbf{37.19}\\
\bottomrule
\end{tabular}
\end{table*}

\begin{table}
\centering
\caption{Quantitative comparison of 3D generation performance across semantic and structural metrics.}
\label{qua3d}
\small 
\setlength{\tabcolsep}{4pt} 
\begin{tabular}{c c c | c c}
\toprule
& \multicolumn{2}{c}{\textbf{Semantic-Level}} & \multicolumn{2}{c}{\textbf{Structure-Level}} \\
{\textbf{Model}} & \textbf{CLIP Similarity$\uparrow$} & \textbf{LPIPS$\downarrow$} & \textbf{CD$\downarrow$} & \textbf{EMD$\downarrow$}\\
\midrule
\textbf{DreamGaussian} & 0.4207 & 0.6675 & 18.43 & 35.59 \\
\textbf{GSGEN} & 0.4834 & 0.6954 & 6.72 & 24.18 \\
\textbf{GraphDreamer} & 0.602 & 0.689 & 10.68 & 14.55 \\
\textbf{Ours} & \textbf{0.701} & \textbf{0.664} & \textbf{4.66} & \textbf{10.93}\\
\bottomrule
\end{tabular}
\end{table}

\section{Experiments}
\subsection{Settings of Experiments}
\subsubsection*{\textbf{Dataset}}
We demonstrate the effectiveness of our Mind2Matter framework using the EEG-Image Dataset \cite{spampinato2017deep}. This dataset comprises EEG recordings collected from six subjects who were presented with a total of 2,000 object images. The visual stimuli were selected from a subset of the widely-used ImageNet dataset \cite{russakovsky2015imagenet}, consisting of 40 categories with 50 easily recognizable images per category. EEG signals were captured using a 128-electrode system at a sampling rate of 1,000 Hz, with each image displayed for 500 ms. For analysis, we utilized 440 ms signal segments to eliminate potential interference from previously displayed images, while excluding the initial 40 ms of each recorded EEG sequence. The dataset covers diverse image categories including animals, vehicles, fruits, etc. Following a standard data partitioning strategy \cite{kavasidis2017brain2image}, we divided the dataset into training, validation, and test subsets with an 8:1:1 ratio.

\subsubsection*{\textbf{Implementation Details}}
Prior to training, all EEG signals were filtered within the frequency range of 55 to 95 Hz. For the EEG encoder, we employed five convolutional layers in the temporal block with progressively increasing kernel dilation values, alongside four convolutional layers in the spatial block. The model parameters were updated using the Adam optimizer, with a learning rate of $2 \times 10^{-5}$, a batch size of 8, and training conducted over 100 epochs. For fine-tuning the LLM, we utilized mistralai/Mistral-7B-Instruct-v0.3 to generate text, adopting the same training hyperparameters as the encoder. For the optimization of 3D Gaussians, we employed MVDream \cite{shi2023mvdream} as the multiview diffusion model, with a guidance scale of 50. The guidance scale for ControlNet was set to 100. All experiments were conducted on a single  RTX 4090 GPU.

\subsubsection*{\textbf{Evaluation Metrics}}
To evaluate the performance of our model, we employ metrics that assess both the quality of the intermediate text generation and the fidelity of the final 3D reconstruction. For the generated text, we utilize standard natural language generation metrics including ROUGE \cite{lin2004rouge} to measure lexical overlap between generated and reference descriptions, BLEU \cite{papineni2002bleu} for n-gram precision evaluation across multiple phrase lengths, and BERTScore \cite{zhang2019bertscore}  to quantify semantic similarity through contextual embeddings. 
For the generated 3D objects, we assess quality at both semantic and geometric levels. Semantically, we use CLIP Similarity \cite{radford2021learning} and Learned Perceptual Image Patch Similarity (LPIPS) \cite{zhang2018unreasonable} to evaluate the perceptual similarity between rendered images and visual stimulus. Geometrically, we apply Chamfer Distance (CD) \cite{fan2017point} and Earth Mover’s Distance (EMD) \cite{ling2007shape} to quantify surface precision and global structural fidelity, respectively. These metrics collectively provide a comprehensive evaluation of our model’s ability to translate EEG signals into accurate and meaningful 3D reconstructions.

\subsection{Quantitative Comparison}
\subsubsection*{\textbf{EEG-to-Text Evaluation:}}
The quantitative results in Table.\ref{quetg} demonstrate the effectiveness of the proposed EEG encoder in the EEG-to-text generation task. Our model consistently outperforms the baseline encoders---Conformer, EEGNet, and DeepNet---across all evaluated metrics. Specifically, our approach achieves the highest ROUGE-1 score of 33.12\%, surpassing Conformer (30.98\%), EEGNet (27.87\%), and DeepNet (23.92\%), indicating superior overlap with reference descriptions. In terms of BLEU scores, our model excels with 34.21\% (B-1), 17.31\% (B-2), 10.06\% (B-3), and 7.62\% (B-4), reflecting better n-gram precision compared to the baselines, particularly on higher-order n-grams (B-3 and B-4), where Conformer scores 15.32\% and 9.02\%, respectively. Furthermore, our model achieves a BERTScore of 37.19\%, outperforming Conformer (35.41\%), EEGNet (30.37\%), and DeepNet (27.16\%), indicating stronger semantic alignment with ground-truth descriptions. These results demonstrate the proposed EEG encoder’s ability to extract robust spatiotemporal features from noisy EEG signals, enabling accurate and semantically coherent text generation.

\begin{figure*}[t]
  \centering
  \includegraphics[width=0.9\textwidth]{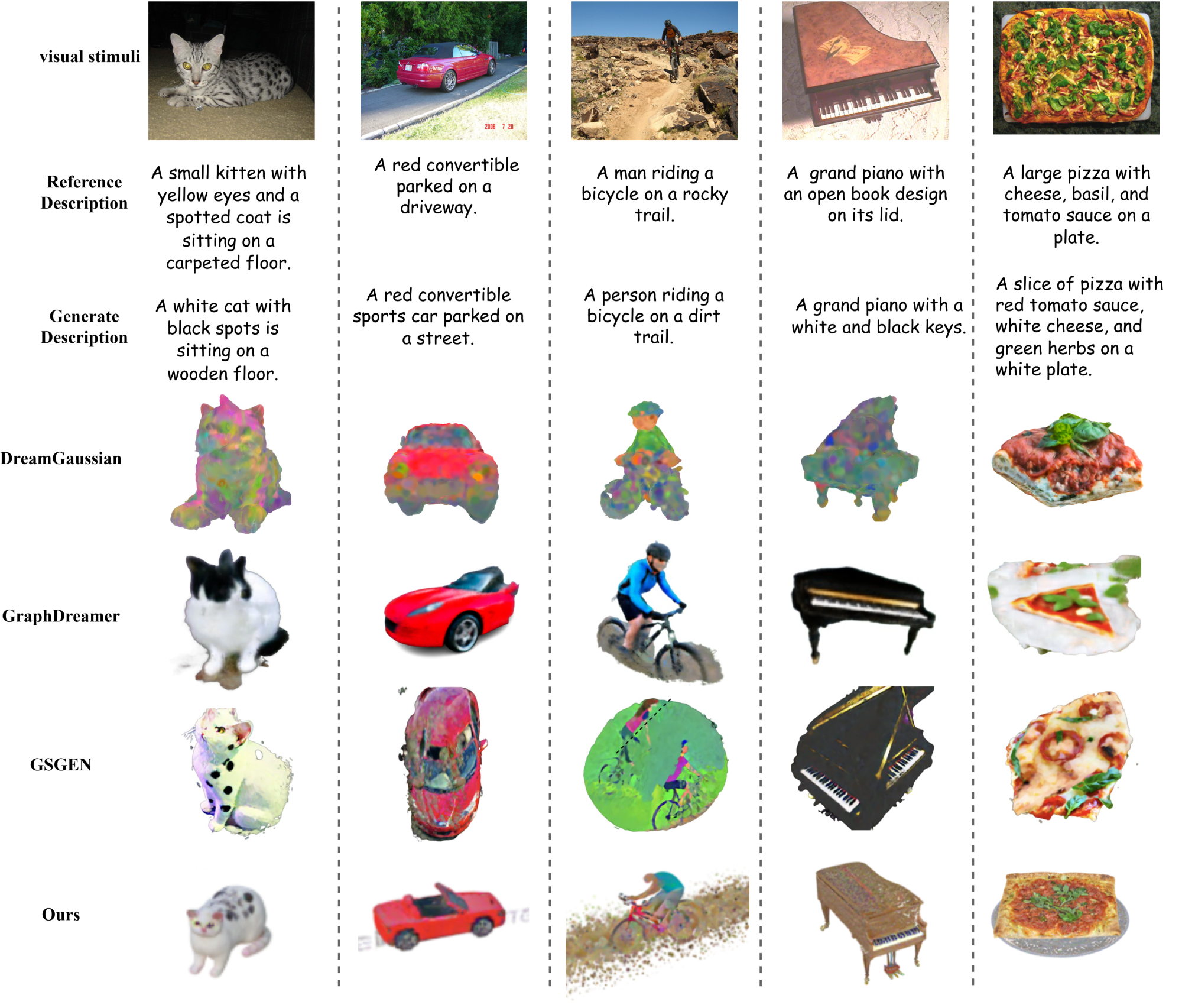} 
  \caption{Qualitative comparison of EEG-to-3D generation results. Comparison between Mind2Matter and baseline methods (DreamGaussian, GraphDreamer, and GSGEN) on four representative visual stimuli. Each row shows: (top) the original visual stimulus with reference description; (middle) the text description generated from EEG signals; (bottom) the final 3D reconstruction results from different methods. Mind2Matter demonstrates superior performance in both semantic preservation and geometric fidelity, while baseline methods exhibit various artifacts such as duplicated components or structural distortions.}
  \label{qual1}
  \Description{}
\end{figure*}

\subsubsection*{\textbf{Text-to-3D Evaluation:}}
Table.\ref{pipline} presents the averaged metrics at both the Semantic and Structural Levels for the 3D reconstruction task from EEG signals across all evaluated methods. The proposed Mind2Matter framework achieves a CLIP Similarity of 0.701, an LPIPS of 0.664, a Chamfer Distance (CD) of 4.66, and an Earth Mover’s Distance (EMD) of 10.93, compared to DreamGaussian with scores of 0.4207, 0.6675, 18.43, and 35.59, and GSGEN with scores of 0.4834, 0.6954, 6.72, and 24.18, respectively. These results indicate that Mind2Matter consistently outperforms the baselines in semantic alignment, as evidenced by the 67\% and 45\% higher CLIP Similarity scores compared to DreamGaussian and GSGEN, respectively, while also achieving better structural accuracy with CD and EMD values that are approximately 75\% and 63\% lower than DreamGaussian, and 31\% and 55\% lower than GSGEN. The LPIPS score of Mind2Matter, although slightly lower than DreamGaussian by 0.004, remains competitive, suggesting that the perceptual quality of the reconstructed 3D models is comparable across methods, with Mind2Matter excelling in capturing both semantic and geometric details.

\subsection{Qualitative Comparison}
We present a qualitative comparison of EEG-to-3D reconstruction results against baseline methods, including DreamGaussian \cite{tang2023dreamgaussian}, GSGEN \cite{chen2024text}, and GraphDreamer \cite{gao2024graphdreamer}. Fig.\ref{qual1} illustrates the reconstruction outcomes for four representative visual stimuli, each accompanied by the original stimulus, the EEG-derived textual description, and the 3D models generated by each method. In the EEG-to-text translation stage, Mind2Matter effectively translates EEG signals into meaningful descriptions. For instance, the original stimulus "A small kitten with yellow eyes and a spotted coat is sitting on a carpeted floor" is accurately interpreted as "A white cat with black spots is sitting on a wooden floor," preserving all key semantic elements despite minor variations in color and texture attributes. This demonstrates Mind2Matter's ability to extract semantic features from noisy EEG signals.

\begin{table*}[ht]
\centering
\caption{Ablation study of key components in the EEG-to-text framework. Performance is evaluated using ROUGE-1 (Recall/R, Precision/P, F1-score/F), BLEU (B-1 to B-4), and BERTScore (R/P/F) metrics. Bold values indicate the full model's performance.}
\label{quaas}
\begin{tabular}{c c c c c c c c c c c}
\toprule
% \textbf{Model} & \textbf{R-1} & \textbf{R-2} & \textbf{R-L} & \textbf{B-1} & \textbf{B-2} & \textbf{B-3} & \textbf{B-4} & \textbf{WER (\%)$\downarrow$} \\
& \multicolumn{3}{c}{\textbf{ROUGE-1(\%)$\uparrow$}} & \multicolumn{4}{c}{\textbf{BLEU(\%)$\uparrow$}} & 
\multicolumn{3}{c}{\textbf{BERT(\%)$\uparrow$}}\\
{\textbf{Model}} & \textbf{R} & \textbf{P} & \textbf{F} & \textbf{B-1} & \textbf{B-2} & \textbf{B-3} & \textbf{B-4} &
\textbf{R} & \textbf{P} & \textbf{F}\\
\midrule
\textbf{w/o GA} &30.93 & 34.48 & 31.75 & 32.28 & 15.16 & 8.83 & 5.98 & 35.24 & 35.34 & 35.13 \\
\textbf{w/o CAML} & 31.39 & 32.13 & 30.85 & 32.16 & 14.46 & 8.22 & 5.62 & 34.44 & 33.29 & 33.88\\
\textbf{w/o Label} & {28.27} & {27.87} & {27.35} & {29.17} & {11.8} & {6.29} & {4.06} & {31.23} & {28.52} & {31.23}\\
\textbf{ALL} & \textbf{33.12} & \textbf{37.38} & \textbf{34.21} & \textbf{34.33} & \textbf{17.31} & \textbf{10.6} & \textbf{7.62} & \textbf{36.84} & \textbf{37.54} & \textbf{37.19}\\
\bottomrule
\end{tabular}
\end{table*}

In the text-to-3D generation phase, our method consistently produces 3D reconstructions that closely capture the semantic intent and spatial details conveyed. In contrast, baseline methods reveal various limitations. DreamGaussian struggles with structural coherence, often generating distorted models. GraphDreamer and GSGEN generate recognizable but structurally flawed results (e.g., two-faced cat, dual-keyboard piano). These errors stem from their reliance on unstable generative processes that struggle to maintain structural coherence during 3D synthesis. Mind2Matter surpasses the baselines by generating 3D scenes with fewer artifacts, precise object details, and superior visual realism. These qualitative results highlight the framework’s ability to effectively translate noisy EEG signals into coherent and detailed 3D representations.

\subsection{Compare with EEG-Image-3D}
\begin{figure}[ht]
  \centering
  \includegraphics[width=0.9\linewidth]{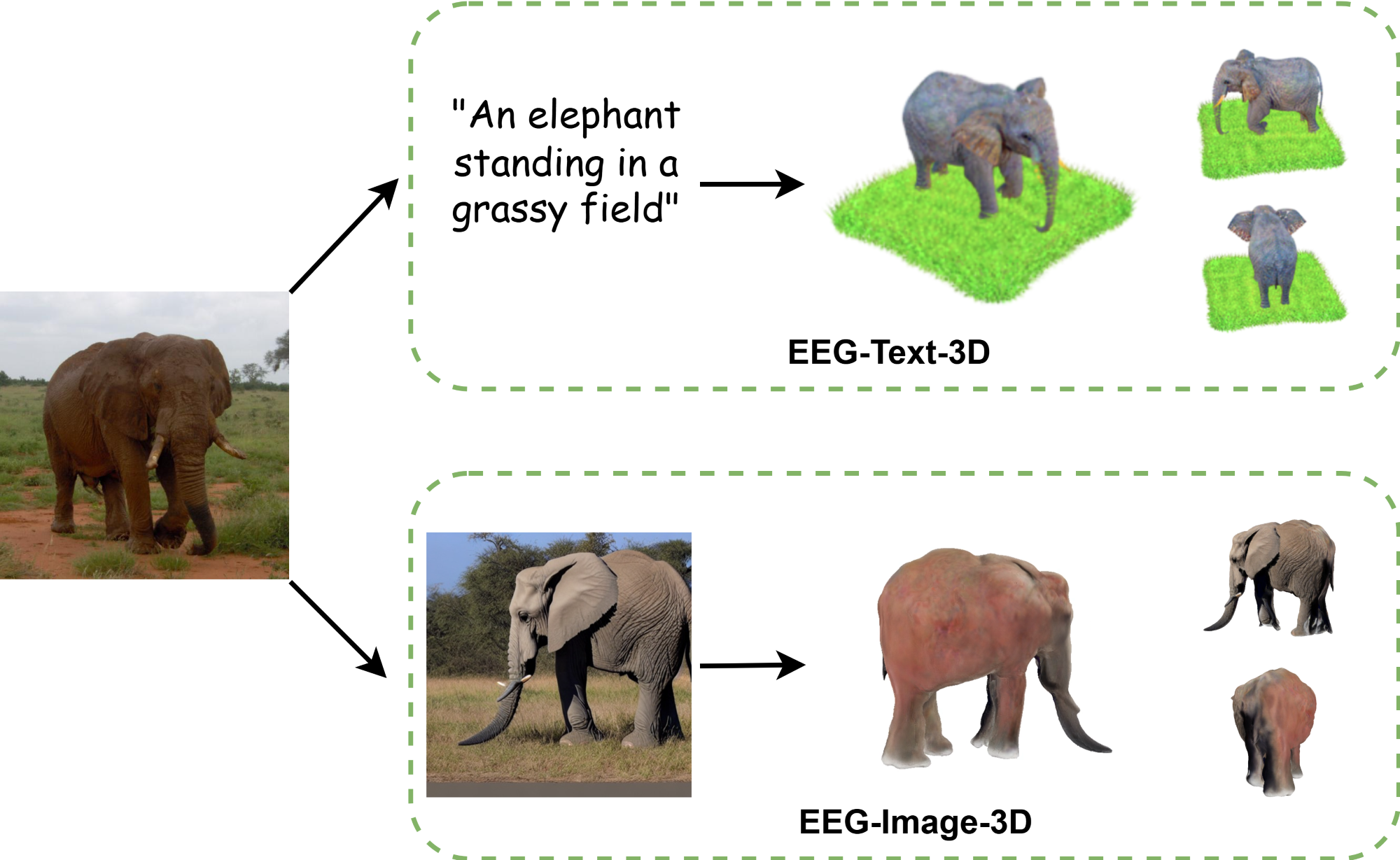}
  \caption{Comparison of 3D Reconstruction Results for EEG-Text-3D and EEG-Image-3D.Top: EEG-Image-3D results show good frontal view but distorted side views. Bottom: Our EEG-Text-3D method maintains consistent quality from all angles.}
  \Description{}
  \label{qual2}
\end{figure}

We compare the proposed EEG-Text-3D pipeline in the Mind2Matter framework with EEG-Image-3D, an approach that reconstructs 3D models by first generating 2D images from EEG signals and then converting these images into 3D representations. Fig.\ref{qual2} presents the results for a stimulus described as "an elephant standing in a grassy field," displaying reconstructions from multiple viewpoints. EEG-Text-3D produces a consistent 3D model with well-defined features—such as the elephant’s trunk, ears, and the grassy field—across frontal, side, and rear views, benefiting from the semantic and spatial guidance provided by the intermediate textual descriptions for a realistic and balanced representation.

In contrast, EEG-Image-3D exhibits notable inconsistencies across perspectives. While the frontal view of the elephant appears visually acceptable, the side and rear views reveal distortions, such as an incomplete trunk and a poorly textured grassy field, due to the limitations of reconstructing 2D images from noisy EEG signals and errors in the subsequent image-to-3D conversion. This comparison highlights EEG-Text-3D’s superiority in achieving robust, view-consistent 3D reconstructions through the use of textual representations.

\subsection{Ablation Study}
We conducted an ablation study by systematically removing or modifying critical modules in the EEG-to-text generation stage. As shown in Table.\ref{quaas}, we evaluated the impact of the Graph Attention module, the Contrastive-Adaptive Margin Loss, and the inclusion of ImageNet label supervision during inference.

\subsubsection*{\textbf{Graph Attention Module:}}Removing the GA module, which models inter-electrode relationships in the EEG encoder, leads to a 2.46\% drop in ROUGE-1 F1-score and a 1.64\% decrease in BLEU-4. This degradation highlights the GA module’s role in capturing relational dependencies among EEG channels, enabling the encoder to extract more robust spatiotemporal features critical for accurate semantic decoding.

\subsubsection*{\textbf{Contrastive-Adaptive Margin Loss:}}
When replacing our Contrastive Adaptive Margin Loss with a standard contrastive loss, performance degrades significantly (3.36\% reduction in ROUGE-1 F1, 2.0\% drop in BLEU-4). These reductions underscore CAML’s importance in adaptively adjusting the discrimination boundary between EEG and visual embeddings, enhancing the model’s ability to align noisy neural signals with semantic representations.

\subsubsection*{\textbf{Label Supervision:}}
Omitting object labels during inference causes the most severe performance drop (6.86\% in ROUGE-1 F1, 3.56\% in BLEU-4). This highlights the importance of semantic supervision in guiding the language model to generate accurate descriptions. Without label conditioning, the model struggles to disambiguate noisy EEG patterns, leading to vague or incorrect text outputs.

\section{Conclusion}
In this paper, we present Mind2Matter, an innovative framework that pioneers EEG-based 3D object reconstruction. Technologically, we propose a novel two-stage approach designed to decode brain activity into tangible 3D models. In the first stage, our EEG encoder proficiently extracts spatiotemporal features from neural signals. The second stage employs a fine-tuned language model and 3D Gaussian splatting with layout control to transform these features into detailed 3D scenes. Evaluated on the EEG-Image Dataset, our model demonstrates effectiveness in capturing semantic and geometric properties, setting a new benchmark in EEG-driven BCI research. Unlike costly fMRI methods, Mind2Matter leverages EEG’s affordability and real-time potential, proving the feasibility of this task. This work not only proves the feasibility of translating noisy, low-resolution EEG signals into structured 3D representations but also provides a foundation for future advancements in neural decoding and 3D generation.

\bibliographystyle{ACM-Reference-Format}
\bibliography{sample-base}

\appendix

\section{Details of Graph Attention}
The Graph Attention Module serves as a critical component in our EEG encoder, designed to capture spatial relationships among electrodes by representing their interactions as a graph-based structure. We formulate the 128-channel EEG montage as a graph $\mathcal{G} = (\mathcal{V}, \mathcal{E})$, where nodes $v_i \in \mathcal{V}$ represent electrodes and edges $e_{ij} \in \mathcal{E}$ encode their functional connectivity. This connectivity criterion ensures that the graph reflects the spatial proximity of electrodes, enabling the module to focus on local inter-electrode relationships while maintaining computational efficiency.

The GA module employs a multi-head attention mechanism comprising 4 attention heads to model diverse spatial dependencies among electrodes. Each head processes the input features, which are 64-dimensional channel-wise representations extracted from the EEG signals after an initial temporal convolution block. These features are transformed into a 128-dimensional space through a linear transformation, followed by a LeakyReLU activation function with a negative slope of 0.2 to introduce non-linearity and enhance gradient flow. The attention coefficients, which quantify the relative importance of neighboring electrodes, are computed as:
\begin{equation}
\alpha_{i,j} = \frac{\exp(\mathbf{a}^T \cdot \text{LeakyReLU}(\mathbf{W}[\mathbf{h}_i \| \mathbf{h}_j]))}{\sum_{k \in \mathcal{N}_i \cup \{i\}} \exp(\mathbf{a}^T \cdot \text{LeakyReLU}(\mathbf{W}[\mathbf{h}_i \| \mathbf{h}_k]))}
\end{equation}
where $\mathbf{h}_i$ and $\mathbf{h}_j \in \mathbb{R}^{64}$ represent the feature vectors of nodes $i$ and $j$, $\mathbf{W} \in \mathbb{R}^{128 \times 64}$ is the weight matrix of the linear transformation, $\mathbf{a} \in \mathbb{R}^{256}$ is a learnable attention vector, and $||$ denotes concatenation. The LeakyReLU activation ensures numerical stability during attention computation. For each head, the updated representation of a node is a weighted sum of the transformed features of its neighbors, where the weights are the attention coefficients. The outputs from all 4 heads are concatenated to produce a final 512-dimensional representation per node (4 heads $\times$ 128 dimensions), capturing a rich set of spatial interactions.

The GA module is strategically placed after the temporal convolution block in the EEG encoder, allowing it to operate on temporally refined features while focusing on spatial patterns. By dynamically weighting the contributions of neighboring electrodes, the module emphasizes spatially significant relationships, such as those between adjacent electrodes that may exhibit correlated neural activity during visual stimulus processing. This design choice enhances the EEG encoder’s ability to extract robust spatiotemporal features, which are critical for aligning EEG embeddings with textual and visual representations in the Mind2Matter pipeline. The use of multi-head attention further ensures that the module captures a diverse range of spatial dependencies, mitigating the risk of overlooking subtle but important inter-electrode interactions. Consequently, the GA module significantly contributes to the framework’s overall performance in EEG-to-text generation, as demonstrated in the ablation study (Section 4.4), where its removal led to a notable decline in semantic accuracy.

\begin{table*}
\centering
\caption{Quantitative comparison of EEG-to-text generation performance across different LLMs. Results are reported in ROUGE-1 (Recall/R, Precision/P, F1-score/F), BLEU (B-1 to B-4), and BERTScore (R/P/F) metrics. Bold values indicate the best performance.}
\label{subt1}
\begin{tabular}{c c c c c c c c c c c}
\toprule
& \multicolumn{3}{c}{\textbf{ROUGE-1(\%)$\uparrow$}} & \multicolumn{4}{c}{\textbf{BLEU(\%)$\uparrow$}} & 
\multicolumn{3}{c}{\textbf{BERT(\%)$\uparrow$}}\\
{\textbf{Model}} & \textbf{R} & \textbf{P} & \textbf{F} & \textbf{B-1} & \textbf{B-2} & \textbf{B-3} & \textbf{B-4} &
\textbf{R} & \textbf{P} & \textbf{F}\\
\midrule
\textbf{Gemma-3-1b} &25.51 & 28.1 & 26.0 & 27.97 & 10.55 & 5.3 & 7.18 & 36.57 & 36.34 & 36.12 \\
\textbf{Llama-3.1-8B} &32.62 & 36.6 & 33.96 & 33.66 & 17.02 & 9.75 & 5.38 & 35.12 & 35.40 & 35.27\\
\textbf{Qwen2.5-7B} & 30.54 & 31.96 & 28.53 & 30.37 & 13.92 & 8.15 & 5.38 & 35.12 & 35.40 & 35.27\\
\textbf{Mistral-7B-v0.3} & \textbf{33.12} & \textbf{37.38} & \textbf{34.21} & \textbf{34.33} & \textbf{17.31} & \textbf{10.6} & \textbf{7.62} & \textbf{36.84} & \textbf{37.54} & \textbf{37.19}\\
\bottomrule
\end{tabular}
\end{table*}

\section{Details of Large Language Model}
In the EEG-to-text generation stage, we utilize a Large Language Model (LLM) to translate EEG-derived embeddings into coherent textual descriptions. To identify the most suitable LLM for this task, we compare the text generation capabilities of four models: Gemma-3-1B, Llama-3.1-8B, Qwen2.5-7B, and Mistral-7B-v0.3. Each model is fine-tuned on the EEG-Image dataset with ImageNet labels as ground-truth textual descriptions, mapping the EEG encoder’s 512-dimensional output to the model’s input space via a two-layer MLP with ReLU activation. The performance of these LLMs is evaluated using standard natural language generation metrics, with results summarized in Figure \ref{subt1}. Mistral-7B-v0.3 outperforms the other models across all metrics, demonstrating superior semantic accuracy and fluency in generating descriptions. Consequently, we select Mistral-7B-v0.3 as the LLM for the Mind2Matter pipeline.

Mistral-7B-v0.3 leverages a decoder-only transformer architecture with 7 billion parameters, incorporating grouped-query attention to optimize computational efficiency while maintaining high-quality text generation. This design makes it particularly effective for the cross-modal task of mapping EEG features to textual outputs, where preserving semantic content and ensuring fluency are paramount. During fine-tuning, the model is adapted to align EEG embeddings with corresponding textual representations, enabling it to capture the semantic essence of visual stimuli encoded in the EEG signals. Integrated after the EEG encoder, Mistral-7B-v0.3 enhances the pipeline’s ability to produce accurate and meaningful text, effectively bridging the gap between neural signals and natural language in the Mind2Matter framework.

\section{More Results}
\begin{figure*}[t]
  \centering
  \includegraphics[width=0.9\textwidth]{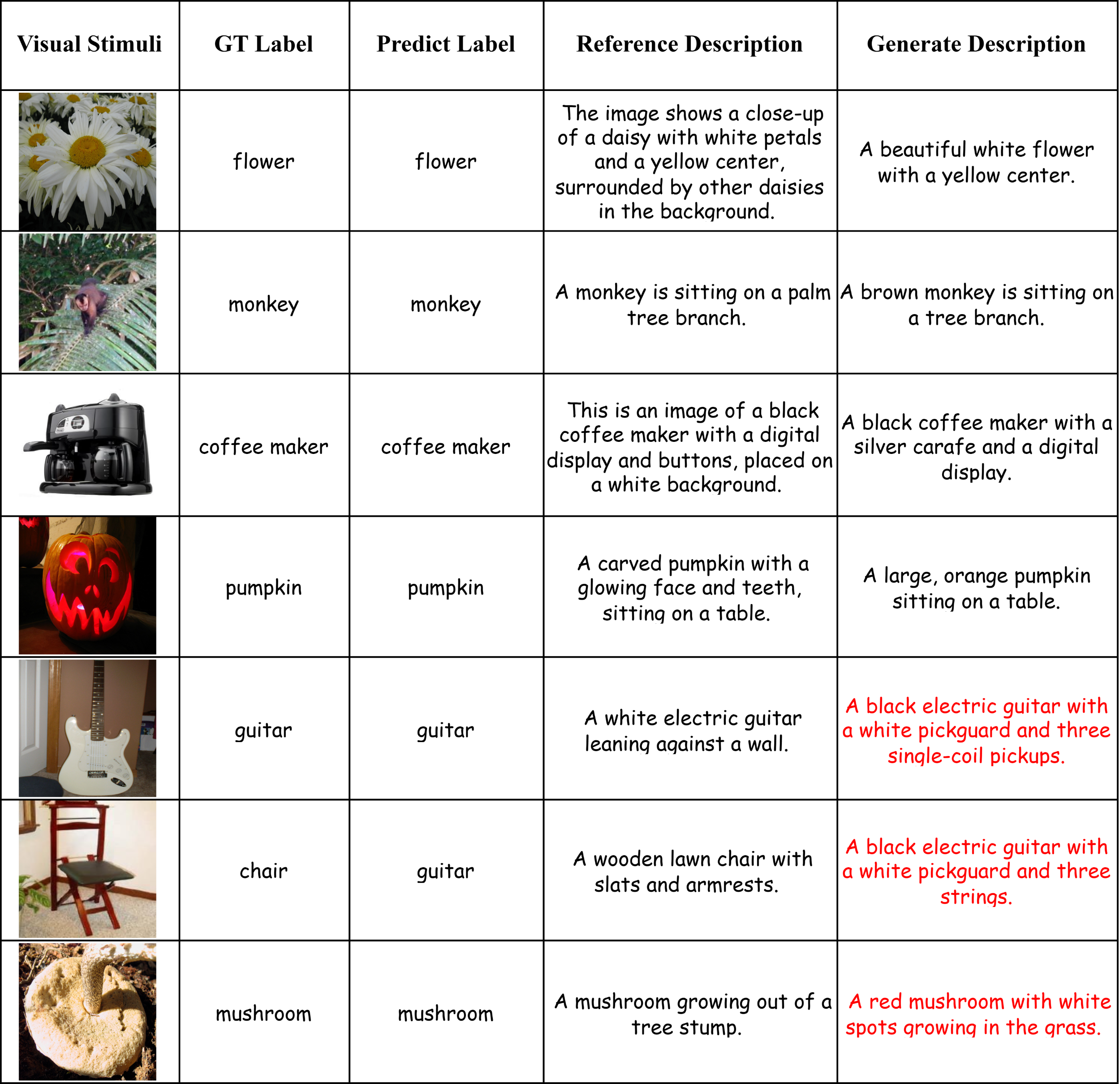} 
  \caption{Additional EEG-to-Text Generation Results.}
  \label{subp1}
\end{figure*}

We present additional results for the EEG-to-text generation task in Figure.\ref{subp1} to further evaluate the Mind2Matter framework. Generated descriptions capture key visual attributes, including color ("white petals," "black coffee maker"), structural details ("yellow center," "digital display"), and spatial context ("sitting on a tree branch"), despite being more concise than reference descriptions. The single misclassification (chair mislabeled as guitar) highlights a known challenge in EEG-based decoding, where objects with similar silhouettes or textures may be confused. In such cases, the model may propagate label errors to the generated text or hallucinate plausible but incorrect details (e.g., describing a chair as having "three strings"). These results illustrate the framework's ability to extract meaningful semantic information from EEG signals while also revealing limitations in fine-grained discrimination, particularly for visually analogous categories. Overall, the generated descriptions provide sufficient accuracy and detail to support downstream 3D reconstruction, validating the efficacy of our approach in bridging neural activity and language-based scene representation.

\section{Failure case}
Despite demonstrating promising EEG-to-3D reconstruction capabilities, our framework exhibits several characteristic failure modes that reveal fundamental challenges in cross-modal translation. The inherent noise and low spatial resolution of EEG signals occasionally lead to category confusions, particularly for visually similar objects sharing comparable silhouettes or textures (e.g., misclassifying chairs as guitars), which subsequently propagates errors through both the generated descriptions and 3D reconstructions. Spatial relationships in complex scenes sometimes suffer from inaccuracies, with overlapping objects or precise relative positioning being challenging to resolve due to limitations in both neural signal interpretation and layout optimization. Furthermore, while the pipeline captures coarse geometric structures effectively, high-frequency details and intricate textures often become oversimplified during the Gaussian splatting process. These limitations primarily stem from three factors: (1) information loss during the EEG-to-semantic encoding stage, (2) the inherent trade-off between reconstruction fidelity and computational efficiency in 3D Gaussian representations, and (3) current constraints in modeling precise physical interactions between objects. Future improvements could incorporate multi-scale EEG feature extraction to better preserve fine details, integrate physics-aware constraints during scene optimization, and employ hybrid representations that combine the efficiency of Gaussian splatting with neural texture synthesis.

\end{document}